\documentclass{article}
\usepackage{romp}
\usepackage{lipsum}
\usepackage{amsfonts}
\usepackage{graphicx}
\usepackage{epstopdf}
\usepackage{algorithmic}
 \usepackage[usenames,dvipsnames]{pstricks}
 \usepackage{amsopn}
 \usepackage{amsmath}
 \usepackage{cite}
 
\newenvironment{Eqnarray}{\arraycolsep 0.14em\begin{eqnarray}}{\end{eqnarray}}
\newcommand{\E}[1]{\langle#1 \rangle}   

\title{ An Advanced Kinetic Theory for Morphing Continuum with Inner Structures }
\author{ James Chen \\ 2090 Rathbone Hall, Department of Mechanical and Nuclear Engineering,\\ Kansas State University, Manhattan, KS 66506, USA\\ 
e-mail: jmchen@ksu.edu \\[2ex] }

\begin{document}

\maketitle
\begin{abstract}
Advanced kinetic theory with the Boltzmann-Curtiss equation provides a promising tool
for polyatomic gas flows, especially for fluid flows containing inner structures, such as turbulence, polyatomic gas flows and others. Although
a Hamiltonian-based distribution function was proposed for diatomic gas flow, a general distribution
function for the generalized Boltzmann-Curtiss equations and polyatomic gas flow is still out of reach. With
assistance from Boltzmann's entropy principle, a generalized Boltzmann-Curtiss distribution for polyatomic
gas flow is introduced. The corresponding governing equations at equilibrium state are derived and compared with
Eringen's morphing (micropolar) continuum theory derived under the framework of rational continuum thermomechanics.
Although rational continuum thermomechanics has the advantages of mathematical rigor and simplicity, the presented statistical
kinetic theory approach provides a clear physical picture for what the governing equations represent.
\end{abstract}


\section{Introduction}

The Navier-Stokes equations have been extensively used to study fluid
flow physics for several decades.  In the fluid mechanics society, it is
usually believed that the Navier-Stokes equations are derived from
kinetic theory \cite{Huang1963} under the assumption that the system of
monatomic gases is nearly in a Boltzmann distribution (equilibrium).  A
system of monatomic gases in equilibrium defines all the physical
quantities in the Boltzmann distribution.  For example, the Navier-Stokes
equations can be derived from the Boltzmann distribution with a first order
approximation applied to the Boltzmann transport equations.

Alternatively, beginning in the early 1960s,
Eringen \cite{Eringen1971, Eringen1980}, Truesdell
\cite{Truesdell1966, TruesdellRajagopal1999} and others
\cite{Chen2013, ChenLeeLiang2011, CimmelliSellitto2009} applied
rational continuum thermomechanics, with its mathematical rigor,
to investigate continuum theories in solids, fluids, mixtures \cite{Bowen1976,
Eringen1997} and electrodynamics.  The foundation of rational continuum
thermomechanics starts with a set of balance laws, including continuity, linear momentum, angular momentum, and energy conservation equations.
With the assistance of several axioms, such
as objectivity (also known as frame-indifference), memory, etc., and the
Coleman-Noll procedure \cite{ColemanNoll1963} based on the Clausius-Duhem
inequality (entropy principle), constitutive equations for a fluid
can be derived.  The combination of balance laws and constitutive
equations for fluids from this approach can also lead to the Navier-Stokes
equations \cite{Chen2013, Eringen1980, Truesdell1966}.

Most of the complex fluid flows contain structures across multiple spatial length scales and are usually dominated by the subscale motions. For example, the rotational eddies characterizes the turbulence physics at bulk scale. In addition, most gases are either diatomic or polyatomic. At high altitude or during speed flights, the rotation and vibration of the diatomic/polyatomic molecules causes non-equilibrium flow pheonmena extensively. Those flows are known for not being able to be analyzed by the classical continuum mechanics or Navier-Stokes equations. As a result, the continuum theory used to model these flows with subscale or inner structures shall be categorized as morphing continuum.

Although rational continuum thermomechanics provides mathematical rigor to the
theoretical formulation, it leaves the physical meanings of material
constants in resulting fluid equations unexplained.  It is common
for practitioners of the rational continuum thermomechanics
approach to interpret the material constants through experiments.  In
contrast, the kinetic theory approach gives physical meaning to the
quantities in the fluid equations by the means of the collision and distribution
functions.  Due to the lack of mathematical rigor in the kinetic
theory approach, Truesdell commented that all equations derived from
kinetic theory can be considered only as a class of constitutive
equations in rational continuum thermomechanics \cite{Truesdell1984}.
Although this comment may be true, the physics hidden in all proposed
governing equations derived from kinetic theory can be used to intepretate
 the same equations using the rigorous procedure of rational
continuum thermomechanics. Such is the purpose of this paper.

Despite the successes of the Navier-Stokes equations derived by both
kinetic theory and rational continuum thermomechanics, the derivations
have been limited to monatomic gases or rather, volumeless point particles. 
Inspired by the rigor of rational continuum thermomechanics, Eringen was the first to
mathematically formulate the micropolar continuum theory with independent inner structures
\cite{Eringen1964_2, Eringen1964_1}.  This work was later expanded by
Chen et.\ al.\ for nonlinear constitutive equations for fluids with
electromagnetic interactions \cite{ChenLeeLiang2011}. Through the multiscale
formulation coupling macroscale and subscale motions, this morphing
continuum theory successfully reproduces and explains turbulence physics \cite{Alizadeth2011, Ahmadi1975, Peddieson1972, Wonnell2017}, polyatomic gases \cite{Chen2016}
and microfluids \cite{Delhommelle2002, Papautsky1999}
which cannot be explained by the classical
Navier-Stokes theories.  Nevertheless, a disadvantage of the rational
continuum thermomechanics approach is that the physical interpretations
of the material constants in MCT are still unknown. Consequently, researchers
rely on dimension analysis to decipher the underlying physics. Wonnell and Chen
presented a systematic derivation on Eringen's micropolar theory for incompressible
turbulence and found a dimensionless parameter differentiating laminar, transition
and turbulent flows. One could also rely on
advanced kinetic theory to deduce the
physical meanings of these MCT material constants. This would be done in a process similar to the one used to derive the Navier-Stokes equations.  This
newly introduced advanced kinetic theory would go beyond the classical
formulation, which is limited to monatomic gases, and be able to treat
polyatomic gases.

The Boltzmann equation was originally developed for a gas consisting
of point particles (i.e., a monatomic gas) and was based on physical
arguments \cite{Boltzmann1872, Curtiss1992}.  Curtiss later expanded
Boltzmann's original formulation to diatomic gas molecules
\cite{Curtiss1981} and polyatomic gas molecules \cite{Curtiss1992}
to obtain what are known as the Boltzmann-Curtiss equations. Although Eu and Ohr \cite{EuOhr2001} proposed
a Hamiltonian-based  distribution function for diatomic gases using Curtiss' transport equation proposed in 1981 \cite{Curtiss1981},
most studies of the
Boltzmann-Curtiss equation adopt the classic Boltzmann distribution with
approximations \cite{GrmelaLafleur1998, Myong2004}.  This leads to one major question
remaining unanswered; namely, what is the distribution that should be used
for the generalized Boltzmann-Curtiss equation \cite{Curtiss1992}?
In this study, a general Boltzmann-Curtiss distribution function (Appendix \ref{app:Derivation}) that accounts
for the translation and rotation of molecular gases is presented. The corresponding governing equations
at equilibrium state are derived and compared with Eringen's morphing (Micropolar) continuum theory.

Section~\ref{sec:KT} then introduces an advanced kinetic theory with a general distribution function, which
accounts for bulk scale translational velocities and subscale rotational motions,
the Boltzmann-Curtiss distribution, and the resulting transport equations.
Section~\ref{sec:KTMC} shows the pathway to the inviscid version of
morphing continuum theory from advanced kinetic theory (see
Section~\ref{sec:KT}) and then compares it to the set of equations obtained from
rational continuum thermomechanics.
Section~\ref{sec:conc} finally gives a brief discussion about the current study
and the next steps.

\section{Advanced Kinetic Theory} \label{sec:KT}

\subsection{Transport Equations}
The Boltzmann equation and kinetic theory have been considered as the
fundamental equations of hydrodynamics and combine atomic-level physics with statistical
averaging.  It is well known that the Navier-Stokes equations can be
deduced from kinetic theory and the Boltzmann equations using a first
order approximation.  However, it should be emphasized that the original
Bolzmann equation and kinetic theory are based on a monatomic particle
assumption.  Therefore, more ad hoc approximations have to be made for
diatomic or polyatomic molecular gases.  Assuming the dynamics of the collisions
do not depend on the vibrational energy, Curtiss
extended the original formulations for the Boltzmann equation and found
the generalized Boltzmann equation (Boltzmann-Curtiss equation) as
\cite{Curtiss1981, Curtiss1992}
\begin{align}
   &\bar{f}\left(\mathbf{x}, \mathbf{p}, \boldsymbol{\phi}, \mathbf{M}, t\right)=\int
   f\left(\mathbf{x}, \mathbf{p}, \boldsymbol{\phi}, \mathbf{M}, E_{\text{vib}},\tau,
   t\right)dE_{\text{vib}}d\tau\nonumber, \\
   &\left(\frac{\partial}{\partial
   t}+\frac{\mathbf{p}}{m}\frac{\partial}{\partial\mathbf{x}}+\frac{\mathbf{M}}{I}
   \frac{\partial}{\partial\boldsymbol{\phi}}\right)\bar{f}=\sum_\beta\mathbf{Z}_\beta,
\label{eq:BCE}
\end{align}
where $\mathbf{x}$ is the position, $\mathbf{p}$ the linear momentum,
$\boldsymbol{\phi}$ the orientation, $\mathbf{M}$ the angular
momentum, $\mathbf{Z}$ the collision integral integrated over $\beta$ molecules that interact with a given molecule, and $I$ the moment
of inertia.  It should be noted that Curtiss did not provide a
distribution function for the Boltzmann-Curtiss equations.  This
study presents a distribution that works well with the Boltzmann-Curtiss
equations.

\subsection{Boltzmann-Curtiss Distribution Function}

Let $\chi$ be any conserved kinetic variable associated with a molecule
of velocity $\mathbf{v}$ and angular velocity $\boldsymbol{\omega}$
located at $\mathbf{x}$, such that in any collision at $\mathbf{x}$ one
has $\{\mathbf{p}_1, \boldsymbol{\omega}_1, \mathbf{p}_2,
\boldsymbol{\omega}_2\}\rightarrow\{\mathbf{p}'_1,
\boldsymbol{\omega}'_1, \mathbf{p}'_2, \boldsymbol{\omega}'_2\}$ so that
$\chi_1+\chi_2=\chi'_1+\chi'_2$ where $\chi_1=\chi(\mathbf{x}_1,
\mathbf{p}_1, \boldsymbol{\omega}_1)$.  The conservation theorem relevant
to the Boltzmann-Curtiss equation can be obtained by multiplying
Eq.~\ref{eq:BCE} on both sides by $\chi$ and integrating over
$\mathbf{v}'$ and $\boldsymbol{\omega}'$.  The collision term vanishes and
the average value, $\E{A}$, is obtained as
\begin{align}
   \E{A}\equiv\frac{\int d^3\,\mathbf{v}'d^3\boldsymbol{\omega}'A\bar{f}}{\int
   d^3\mathbf{v}'\,d^3\boldsymbol{\omega}'\bar{f}}=\frac{1}{n}\int
   \,d^3\mathbf{v}'\,d^3\boldsymbol{\omega}'A\bar{f} ,
\end{align}
where $n(\mathbf{x}, t)\equiv \int d^3\,\mathbf{v}'
d^3\boldsymbol{\omega}'\bar{f}(\mathbf{x}, \mathbf{p},
\boldsymbol{\omega}, t)$.

Assume the inner structure as a rigid sphere and has no distribution
inside the molecule, i.e.  $\frac{\partial}{\partial\phi_k}=0$.  In the
absence of body force and couple moment, the conservation equation
is
\begin{align}
   &\frac{\partial}{\partial t}\E{n\chi}+\frac{\partial}{\partial x_i}\E{n\chi v_i}-n\E{v_i\frac{\partial\chi}{\partial x_i}}+\frac{\partial}{\partial \phi_i}\E{n\chi\omega_i}-n\E{\omega_i\frac{\partial\chi}{\partial \phi_i}}=0, \nonumber \\
  &\Rightarrow \frac{\partial}{\partial t}\E{n\chi}+\frac{\partial}{\partial x_i}\E{n\chi v_i}-n\E{v_i\frac{\partial\chi}{\partial x_i}}=0
\label{eq:KTB}
\end{align}
and the derivation of the Boltzmann-Curtiss distribution can be found in Appendix \ref{app:Derivation}. The Boltzmann-Curtiss distribution is
\begin{equation}
   \bar{f}\left(\mathbf{x}, \mathbf{v},
   \boldsymbol{\omega}\right) = n\left(\frac{m\sqrt{j}}{2\pi k\theta}
   \right)^{\! 3} \exp\left[-\frac{m(\mathbf{v}'^2+j\boldsymbol{\omega}'^2)}{2k\theta}\right] .
   \label{eq:BCDis}
\end{equation}
The corresponding probability density function has the proper normalization, i.e.,
\begin{equation}
   \int_{-\infty}^{\infty} \left( \frac{m\sqrt{j}}{2\pi k\theta}\right)^{\! 3}
   \exp\left[-\frac{m(\mathbf{v}'^2+j\boldsymbol{\omega}'^2)}{2k\theta}\right]
   d^3\mathbf{v}\,d^3\boldsymbol{\omega}=1.
\end{equation}
Here $m$ is the mass of the molecule, $j$ is the microinertia for the inner structure
and $k$ is the Boltzmann constant. One can understand microinertia through the concept of moment of inertia. Microinertia is a measurement of the resistance of the internal structure to changes to its rotation and can be defined as
 $$j\equiv2\E{r_mr_m}=2\frac{\int_{\Delta v'}\rho'r_mr_mdv'}{\int_{\Delta v'}\rho'd_v'},$$
where $r_m$ is the local coordinate from the centroid of the internal structure and $\Delta v'$ is the volume of the inner structure.

It should be noted that following the equipartition of energy,
$E_{\text{int}}=\frac{1}{2}N_{\text{DOF}}nk\theta$, the internal energy
can be shown to be
\begin{align}
   E_{\text{int}}=\frac{6}{2}nk\theta=\frac{1}{2}m\left(\E{\mathbf{v}'^2}+
  j\E{\boldsymbol{\omega}'^2}\right) ,
\end{align}
where $N_{\text{DOF}}$ is the number of the degrees of freedom in a
polyatomic molecule and $n$ is the number of the molecules in the
system.  The velocity $v_i$ and the angular velocity $\omega_i$ can
be decomposed as $v_i=U_i+v'_i$ and $\omega_i=W_i+ \omega'_i$.

\subsection{Kinetic Variables}  \label{sec:KV}

The equations for fluid dynamics involving particles with inner structure
can be derived by calculating the moments of the Boltzmann-Curtiss equations
for quantities that are conserved in collisions of the molecules.  There
are four conserved quantities $\chi$, i.e., the mass $\chi_1=m$, the
total linear momentum $\chi_2=mv^*_i=m(v_i+e_{ijk}\omega_jr_k)$, the
angular momentum $\chi_3=mj\omega_m$ and the total energy
$\chi_4=\frac{1}{2}m(v_iv_i+j\omega_i\omega_j)$.

\subsubsection{Balance Law of Mass}
If $\chi_1=m$ is chosen and inserted into Eq.~\ref{eq:KTB}, one obtains
\begin{equation}
   \frac{\partial}{\partial t}\E{nm}+\frac{\partial}{\partial x_i}\E{nmv_i}=0 ,
\end{equation}
With $nm=\rho$ and $\E{v_i}=U_i$, the continuity equation is obtained as
\begin{equation}
   \frac{\partial}{\partial t}\rho+\frac{\partial}{\partial x_i}\left(\rho U_i\right)=0.
\end{equation}

\subsubsection{Balance Law of Linear Momentum}
If $\chi_2=mv^*_i=m(v_i+e_{ijk}\omega_jr_k)$, Eq.~\ref{eq:KTB} becomes
\begin{Eqnarray}
   \frac{\partial}{\partial t}\E{nmv_j}+\frac{\partial}{\partial
   t}(nme_{jmn}\omega_mr_n) &+& \frac{\partial}{\partial
   x_i}\E{nmv_iv_j} \nonumber \\
   &+& \frac{\partial}{\partial x_i}\E{nme_{jmn}\omega_mr_nv_i}=0,
\end{Eqnarray}%
or
\begin{equation}
   \frac{\partial}{\partial t}\left(\rho U_j\right)+\frac{\partial}{\partial
   x_i}\left(\rho U_iU_j\right)=-\frac{\partial}{\partial x_i}\left(\rho\E{v'_iv'_j}+\rho
   e_{jmn}r_n\E{v'_i\omega'_m}\right) .
\end{equation}
Here $U^*_j=U_j+e_{jmn}W_mr_n$ is the total averaged velocity with the
fluctuating velocity integrated over the volume of the molecule being
zero, i.e.,  $\int e_{jmn}W_mr_nd^3\mathbf{v}'d^3\boldsymbol{\omega}'=0$.
The quantity $t^{\text{Boltzmann}}_{ij}=\rho\E{v'_iv'_j}$ is the stress similar to
that in the classical Boltzmann equation and
$t^{\text{Curtiss}}_{ij}=\rho e_{jmn}r_n\E{v'_i\omega'_m}$ is the
asymmetric part of the stress due to the rotation discussed by Curtiss
\cite{Curtiss1992}, Eringen \cite{ArimanCakmak1967, Eringen1964_1},
Stokes \cite{Stokes1966} and others \cite{ChenLeeLiang2011}.  It
should be noticed that the integral for the symmetric part of the stress,
$t^{\text{Boltzmann}}_{ij}$, is different from the one in the classical
Boltzmann equation due to the distribution function used; however, the
derivation still yields the gas pressure and the classical
symmetric stress tensor at Boltzmann-Curtiss distribution, namely
\begin{Eqnarray}
   t^{\text{Boltzmann}}_{ij}&=&p\delta_{ij}-t^{\text{Boltzmann,viscous}}_{ij}
   =\rho\E{v_i'v_j'} \nonumber \\
   &=& \frac{\rho}{n}\int \!\!\!\int n v_i'v_j'\left(\frac{m(j)^\frac{1}{2}}{2\pi
   k\theta}\right)^3e^{-\frac{m\left(\mathbf{v'}^2+j\boldsymbol{\omega'}^2\right)}{2k\theta}}d^3
   \mathbf{v}'\,d^3\boldsymbol{\omega}'.
\end{Eqnarray}%
It is also straightforward to prove that
$\rho\E{v_x'v_x'}=\rho\E{v_y'v_y'}=\rho\E{v_z'v_z'}=nk\theta$ and define the gas
pressure as $p\equiv\frac{1}{3}\rho\E{\mathbf{v'}^2}$ when the system is at the Boltzmann-Curtiss distribution.

\subsubsection{Balance Law of Angular Momentum}

For a rotating body, one of the important quantities is angular
momentum.  Therefore, the third kinetic variable is
$\chi_3=mj\omega_m$, where $j$ is the
inertia of the subscale sturcture.  With the substitution of $\chi_3$, Eq.~\ref{eq:KTB} becomes
\[
   \frac{\partial}{\partial t}\E{nmj\omega_m}+\frac{\partial}{\partial
     x_i}\E{nmj\omega_mv_i}=0
\]
or
\begin{equation}
   \frac{\partial}{\partial t}\left(\rho jW_m\right)+\frac{\partial}{\partial
     x_i}\left(\rho jW_mU_i\right)=-\frac{\partial}{\partial x_i}\left(\rho
    j\E{\omega'_mv'_i}\right) ,
\end{equation}
where the right-hand term, $m_{im}=\rho j\E{\omega'_mv'_i}$,
is the combination of the moment stress due to
rotation and the asymmetric part of the Cauchy stress,
$t^{\text{Curtiss}}_{ij}=\rho e_{jmn}r_n\E{v'_i\omega'_m}$.

\subsubsection{Balance Law of Energy}

In classical kinetic theory, the internal energy density is defined
as $e=\frac{1}{2}\E{v'_mv'_m}$; however, because of the rotating effect of
the molecule, the internal energy density should be re-defined as
$e=\frac{1}{2}(\E{v'_mv'_m}+j\E{\omega'_m\omega'_n})$.  Similarly, the total energy is classically defined as $\frac{1}{2}m(v_mv_m)$, but due to the rotation effects of the molecule it should now be defined as $\frac{1}{2}m(v_mv_m+j\omega_m\omega_m)$. This can also be stated as $\frac{1}{2}m(e+U_mU_m+jW_mW_m)$. After inserting
the last kinetic variable, i.e., total energy, $\chi_4=\frac{1}{2}m(v_mv_m+j\omega_m\omega_m)$, Eq.~\ref{eq:KTB} is now
\begin{align}
   \frac{\partial}{\partial t}\left(\rho e\right)&+\frac{\partial}{\partial x_i}\left(\rho
     eU_i\right)\nonumber \\
   &=U_m\frac{\partial}{\partial
     x_i}\left(\rho e_{mjn}r_n\E{v'_i\omega'_j}\right)-\rho\E{v_i'v_m'}\frac{\partial U_m}{\partial x_i}\nonumber \\
    &-\frac{\partial}{\partial
     x_i}\left(\frac{1}{2}\rho\left(\E{v'_mv'_mv'_i}+j\E{\omega'_m\omega'_nv'_i}\right)\right)-\rho j\E{\omega_m'v_i'}\frac{\partial W_m}{\partial x_i},
\end{align}
or
\begin{align}
   \frac{\partial}{\partial t}\left(\rho e\right)&+\frac{\partial}{\partial x_i}\left(\rho
     eU_i\right)\nonumber \\
   &=t_{im,i}^{\text{Curtiss}}U_m-t_{im}^{\text{Boltzmann}}U_{m,i}-q_{i,i}-m_{im}W_{m,i}.
\end{align}
It should be noted that because of the molecular
rotation, the heat flux density has an additional term involving
rotation and is now defined as
$q_i=\frac{1}{2}\rho(\E{v'_mv'_mv'_i}+j\E{\omega'_m\omega'_nv'_i})$.

\section{Comparison of the Advanced Kinetic and Micropolar Theories for Morphing Continuum}  \label{sec:KTMC}
By introducing the kinetic variables into the Boltzmann-Curtiss
equation, Eq.~\ref{eq:KTB}, the four balance laws can also be found as
\begin{align}
&\frac{\partial}{\partial t}\rho+\frac{\partial}{\partial x_i}\left(\rho U_i\right)=0, \nonumber \\
&\frac{\partial}{\partial t}\left(\rho U_j\right)+\frac{\partial}{\partial x_i}\left(\rho U_iU_j\right)=
   -\frac{\partial}{\partial x_i}\left(\rho\E{v'_iv'_j}+\rho e_{jmn}r_n\E{v'_i\omega'_m}\right),
     \nonumber \\
&\frac{\partial}{\partial t}\left(\rho jW_m\right)+
   \frac{\partial}{\partial x_i}\left(\rho jW_mU_i\right)=-\frac{\partial}{\partial x_i}
     \left(\rho j\E{\omega'_mv'_i}\right)\nonumber \\
&\frac{\partial}{\partial t}\left(\rho e\right)+\frac{\partial}{\partial x_i}\left(\rho
     eU_i\right)=U_m\frac{\partial}{\partial
     x_i}\left(\rho e_{mjn}r_n\E{v'_i\omega'_j}\right)-\rho\E{v_i'v_m'}\frac{\partial U_m}{\partial x_i}\nonumber \\
    &-\frac{\partial}{\partial
     x_i}\left(\frac{1}{2}\rho\left(\E{v'_mv'_mv'_i}+j\E{\omega'_m\omega'_nv'_i}\right)\right)-\rho j\E{\omega_m'v_i'}\frac{\partial W_m}{\partial x_i}
\label{eq:KTB+Eq}
\end{align}

\subsection{The Zero-order Approximation}
If the mean free path is small compared to other characteristic lengths,
the system rapidly comes to a local equilibrium.  In classical
kinetic theory, it is well known that the zero-order approximation
reduces the Boltzmann equations to the governing equations for inviscid
flow (i.e., Euler equations).  Here, the zero-order approximation is applied
to the Boltzmann-Curtiss equations (i.e., to Eq.~\ref{eq:KTB+Eq}).  First
assume that the gas has a local Boltzmann-Curtiss distribution, with
slowly varying temperature, density, velocity and rotation, so the distribution
can be approximated by
\begin{equation}
   \bar{f}(\mathbf{x}, \mathbf{v}, \boldsymbol{\omega})\approx\bar{f}^0\left(\mathbf{x},
   \mathbf{v}, \boldsymbol{\omega}\right)=
   n \left( \frac{m\sqrt{j}}{2\pi k\theta} \right)^3
   \exp\left[-\frac{m(\mathbf{v}'^2+j\boldsymbol{\omega}'^2)}{2k\theta}\right]
\end{equation}
It should be noted that the Boltzmann-Curtiss distribution is not the
exact solution to Eq.~\ref{eq:BCE}; however, it serves as a reasonable
approximation just as the Maxwell-Boltzmann distribution serves as a good
approximation for classical kinetic theory.  In the previous
section, it has already been shown that $\E{v'_iv'_i}=3nk\theta$ and
consequently, the pressure becomes
\begin{align}
   t^{\text{Boltzmann,0}}_{ij}&=P^0_{ij}\nonumber \\
   &=\frac{\rho}{n}\int\!\!\!\int n v_iv_j
   \left( \frac{m\sqrt{j}}{2\pi k\theta} \right)^3
   \exp\left[-\frac{m(\mathbf{v}'^2+j\boldsymbol{\omega}'^2)}{2k\theta}\right]
   d^3\mathbf{v}'d^3\boldsymbol{\omega}'\nonumber \\
   &=nk\theta\delta_{ij} .
\end{align}
In addition, it is straightforward to prove that
\begin{align}
   &t^{\text{Curtiss,0}}_{ij}=\frac{\rho}{n} e_{jmn}r_n\int\!\!\!\int nv'_i\omega'_m
       \left( \frac{m\sqrt{j}}{2\pi k\theta} \right)^3
   \exp\left[-\frac{m(\mathbf{v}'^2+j\boldsymbol{\omega}'^2)}{2k\theta}\right]
       d^3\mathbf{v}'d^3\boldsymbol{\omega}'=0 , \nonumber \\
   &m^0_{ij}=\frac{\rho}{n} j \int\!\!\! \int n\omega'_mv'_i
       \left( \frac{m\sqrt{j}}{2\pi k\theta} \right)^3
   \exp\left[-\frac{m(\mathbf{v}'^2+j\boldsymbol{\omega}'^2)}{2k\theta}\right]
       d^3\mathbf{v}'d^3\boldsymbol{\omega}'=0, \nonumber \\
   &q^0_i=\frac{\rho}{2n}  \int\!\!\! \int n(v'_mv'_mv'_i+j\omega'_m\omega'_nv'_i)
       \left(\frac{m\sqrt{j}}{2\pi k\theta} \right)^3
   \exp\left[-\frac{m(\mathbf{v}'^2+j\boldsymbol{\omega}'^2)}{2k\theta}\right]
        d^3\mathbf{v}'d^3\boldsymbol{\omega}'=0.
\end{align}

With these results, Eqs.~\ref{eq:KTB+Eq} reduce to
\begin{align}
   &\frac{\partial}{\partial t}\rho+\frac{\partial}{\partial x_i}\left(\rho U_i\right)=0,\nonumber \\
   &\frac{\partial}{\partial t}\left(\rho U_j\right)+\frac{\partial}{\partial x_i}\left(\rho U_iU_j\right)=-\frac{\partial}{\partial x_i}\left(n k\theta\delta_{ij}\right),\nonumber \\
   &\frac{\partial}{\partial t}\left(\rho jW_m\right)+\frac{\partial}{\partial x_i}\left(\rho jW_mU_i\right)=0 ,\nonumber \\
   &\frac{\partial}{\partial t}\left(\rho e\right)+\frac{\partial}{\partial x_i}\left(\rho eU_i\right)=-nk\theta U_{i,i}\nonumber \\
\label{eq:KinInv}
\end{align}

\subsection{Comparison with Eringen's Micropolar Continuum Theory}
Inspired by the Cosserat Brothers, Eringen formulated a micropolar continuum theory for fluid flow
containing inner structures \cite{Eringen1964_2, Eringen1964_1}. With the linear constitutive equations, the governing equations are found to be
\begin{Eqnarray}
&& \frac{\partial}{\partial t}\rho+\frac{\partial}{\partial x_i}
       \left(\rho v^{\text{MCT}}_i\right)=0 , \nonumber \\[0.1in]
&&\frac{\partial}{\partial t}(\rho v^{\text{MCT}}_j)+\frac{\partial}{\partial x_i}
    \left(\rho v^{\text{MCT}}_jv^{\text{MCT}}_i\right)= \nonumber \\
    && \qquad \qquad   -p_{,j}+(\lambda+\mu)v^{\text{MCT}}_{m,mj}+
    \left(\mu+\kappa\right)v^{\text{MCT}}_{j,mm}+\kappa e_{jki}\omega^{\text{MCT}}_{i,k} ,
       \nonumber \\[0.1in]
&& \frac{\partial}{\partial t}(\rho i_{jk}\omega^{\text{MCT}}_k)+
     \frac{\partial}{\partial x_i}\left(\rho i_{jk}\omega^{\text{MCT}}_kv^{\text{MCT}}_i\right)=
     \left(\alpha+\beta\right)\omega^{\text{MCT}}_{m,mj}+ \nonumber\\
      && \qquad \qquad \gamma\omega^{\text{MCT}}_{j,mm}+
      \kappa\left(e_{jki}v^{\text{MCT}}_{i,k}-2\omega^{\text{MCT}}_j\right) ,\nonumber \\[0.1in]
&&\frac{\partial}{\partial t}\left(\rho e\right)+\frac{\partial}{\partial x_i}
     \left(\rho e v^{\text{MCT}}_i\right)-t_{ij}\left(v^{\text{MCT}}_{j,i}+
     e_{jim}\omega^{\text{MCT}}_m\right)- \nonumber \\
     && \qquad \qquad m_{lk}\omega^{\text{MCT}}_{k,l}+q_{k,k}=0.
\end{Eqnarray}%

The detailed derivation for Eringen's Micropolar theory can be found in the numerous publications \cite{Chen2013, ChenLeeLiang2011, Eringen1964_2, Eringen1964_1}, and Appendix \ref{app:RCT}. Finally, if all the viscous and dissipative terms are dropped, the MCT
governing equations reduce to
\begin{align}
   &\frac{\partial}{\partial t}\rho+\frac{\partial}{\partial x_i}(\rho v^{\text{MCT}}_i)=0 , \nonumber \\
   &\frac{\partial}{\partial t}\left(\rho v^{\text{MCT}}_j\right)+\frac{\partial}{\partial x_i}\left(\rho v^{\text{MCT}}_jv^{\text{MCT}}_i\right)=-\frac{\partial p}{\partial x_i}, \nonumber \\
   &\frac{\partial}{\partial t}\left(\rho i_{jk}\omega^{\text{MCT}}_k\right)+\frac{\partial}{\partial x_i}\left(\rho i_{jk}\omega^{\text{MCT}}_kv^{\text{MCT}}_i\right)=0 , \nonumber \\
   &\frac{\partial}{\partial t}\left(\rho e\right)+\frac{\partial}{\partial x_i}\left(\rho e v^{\text{MCT}}_i\right)=-p v^{\text{MCT}}_{m,m}.
\label{eq:MCTInv}
\end{align}

In comparing Eqs.~\ref{eq:KinInv} and \ref{eq:MCTInv}, it is seen that
the equations of MCT and advanced kinetic theory are identical!  Kinetic
theory, by nature, is a bottom-up theory involving physical conjectures
and insights while MCT is a top-down theory involving mathematical rigor.
However, both theories lead to the same governing equations for inviscid
fluid flow.
\section{Conclusion}
\label{sec:conc}

With the rapid growth of research in nonequilibrium thermodynamics
of highly compressible flow with high Mach numbers \cite{Klembt2015,
Venugopal2015}, supersonic/hypersonic turbulence \cite{Wonnell2016} and in micro-/nano-fluidics \cite{Norouzi2015}, the traditional kinetic theory, or the classical
continuum theory, is in conflict with some of the physics observed in
the experiments.  One of the major reasons for this conflict is that
most molecules are either polyatomic or diatomic possessing not only
translational but rotational and vibrational energies.
Or the flow itself contains motions from more than
one length scale and the subscale motion strongly
dominates the macroscale motion as in turbulence.
Both the classical
bottom-up kinetic theory and top-down continuum theory consider the
system to be composed of volumeless monatomic molecules so the effects of
the rotational and vibrational effects in polyatomic or diatomic
structures or those of rotational eddies in turbulence
 are ignored and have to be treated in an ad hoc manner.

As shown here, advanced kinetic theory based on the Boltzmann-Curtiss
transport equations can treat the effect of inner structure on the bulk scale
phenomena through additional degrees of freedoms in rotation. Rational
continuum thermodynamics provides the rigorous mathematics needed for a
micropolar continuum theory involving the rotational effect of inner structures.
Instead of distinguishing between the advanced kinetic theory
and micropolar continuum theory, one can impose the mathematical rigor of
rational continuum mechanics on the physical constraints (e.g. entropy
principle) and insights (e.g. molecular collision) of the advanced
kinetic theory for a complex system of fluid flows.

Although neither dissipative nor viscous effects are included in the current
study, an appropriate order approximation on the Boltzmann-Curtiss
distribution can shed new physical insights on the material constants of
the morphing continuum constitutive equations (i.e., on
Eq.~\ref{eq:MCT+Con}).  The integration of micropolar continuum
theory and advanced kinetic theory can serve as an alternative
theoretical tool for nonequilibrium aerothermodynamics for turbulence and high Knudsen
number flows in any system involving translation and rotation of the
particles.

\section*{Acknowledgement}
The author would like to express his gratitude to Dr. Ken Shultis and Dr. Mingjun Wei for their insightful discussions and constructive suggestions. This material is based upon work supported by the Air Force Office of Scientific Research under award number FA9550-17-1-0154.

\appendix
\section{Derivation of the Boltzmann-Curtiss Distribution}  \label{app:Derivation}

The derivation presents a systematic approach for the
Boltzmann-Curtiss distribution with the Boltzmann principle.  Let us assume that a closed system
contains a large number $N$ of distinguishable polyatomic molecules in
thermal equilibrium with its surroundings at a temperature $\theta$ and
under constant volume conditions.  Let $\epsilon_i$ denote the i-th
nondegenerate energy level having $n_i$ molecules, where
$i=0,1,2,3,...,m$.

The number of microstates, $W$, is given by
\begin{equation}
        W=\frac{N!}{\sum_{i=0}^m n_i!} .
\label{eq:microstate}
\end{equation}
It should be noted that $\sum_{i=0}^m n_i= N$.  The connection between
the statistical mechanical description of the system and the bulk
thermodynamic description given by the entropy is expressed in the
Boltzmann principle, namely
\begin{equation}
        S=k \text{ln} W ,
 \label{eq:B_Entropy}
\end{equation}
where k is the well-known Boltzmann constant.

Substitution of Eq. \ref{eq:microstate} into Eq. \ref{eq:B_Entropy} gives
\begin{equation}
        S=k \text{ln} W\nonumber
        =k \text{ln} \left(\frac{N!}{\sum_{i=0}^m n_i!}\right)\nonumber
        =k \left[\text{ln} \left(N!\right) - \sum_{i=0}^m \text{ln} \left(n_i!\right)\right].
\end{equation}
Now suppose that a small amount of energy $\triangle\epsilon$ is added to the
system redistributing the number of molecules in different energy
levels.  Therefore, the change of entropy between the two different states
(before and after the energy injection) can be found and expanded with
a Taylor's Series as
\begin{align}
        \triangle S&=k \sum_{i=0}^m \{\text{ln} \left(n_i!\right) - \text{ln} \left[\left(n_i+\triangle n_i\right)!\right]\} \nonumber \\
        &=k \sum_{i=0}^m \triangle n_i \left[\text{ln} (n_i) - \text{ln} (n_i+\triangle n_i)\right] + O(\triangle n_i^2)\nonumber \\
        &\approx k \sum_{i=0}^m \triangle n_i \text{ln} \left(\frac{n_i}{n_i+\triangle n_i}\right) .
  \label{eq:EntropyDiff}
\end{align}

From the laws of thermodynamics, the connection between the
entropy $S$ and total energy $U$ of the system is $dS=dU/\theta$.
Consequently,
\begin{equation}
        \triangle S= \int dS = \int \frac{dU}{\theta}
        = \frac{1}{\theta} \int dU = \frac{1}{\theta} \triangle U .
\label{eq:internal}
\end{equation}

With the assumption that temperature remains effectively constant, the energy
change $\triangle U$ is the energy $\triangle\epsilon = \frac{1}{2}
m(\triangle v'^2 + j \triangle\omega'^2)$ injected into the original
system.  As a result, Eq.~\ref{eq:internal} becomes
\begin{equation}
        \triangle S = \frac{1}{\theta} \triangle\epsilon
        = \frac{1}{2\theta} m\left(\triangle v'^2 + j\triangle\omega'^2\right) .
\label{eq:InternalEntropy}
\end{equation}
Combine Eqs.~\ref{eq:EntropyDiff} and \ref{eq:InternalEntropy} to obtain

\[
   k \sum_{i=0}^m \triangle n_i \text{ln}
   \left(\frac{n_i}{n_i+\triangle n_i}\right) =
   \frac{1}{2\theta} m\left(\triangle v'^2 + j \triangle\omega'^2\right),
\]
or, equivalently,
\begin{equation}
  \sum_{i=0}^m \left(\frac{n_i + \triangle n_i}{n_i}\right)^{\triangle n_i} =
   \exp\left[-\frac{1}{2k\theta}m\left(\triangle v'^2 + j\triangle\omega'^2\right)\right] .
\label{eq:FinalBCD}
\end{equation}
Equation \ref{eq:FinalBCD} is the Boltzmann-Curtiss distribution (c.f. eq. \ref{eq:BCDis}) used for the Boltzmann$\textendash$Curtiss transport
equations and is the basis of the morphing continuum theory at
equilibrium.

\section{Rational Continuum Thermomechanics for Micropolar Theory}  \label{app:RCT}
A morphing continuum is a collection of continuously distributed,
oriented, finite-size particles that can rotate.  A material point $P$
in the reference frame is identified by a position and three directors
attached to the material point.

The motion, at time $t$, carries the finite-size particle to a spatial
point and rotates the three directors to a new direction.  Thus, such a motion
is similar to the motion of a liquid molecule approximated as a sphere.
It has not only translational velocity, but also
self-spinning gyration on its own axis.  These motions and their inverse
motions for the morphing continuum can be described as
\begin{align}
   x_k&=x_k\left(X_K, t\right) \quad & X_K&=X_K\left(x_k, t\right)\nonumber \\
   \xi_k&=\chi_{kK}\left(X_K, t\right)\Xi_K	\quad & \Xi_K&=\bar{\chi}_{Kk}\xi_k\nonumber \\
   K&=1, 2, 3 \quad &  k&=1, 2, 3
\end{align}
and
\begin{align}\label{eq:inverse}
   \chi_{kK}\chi_{lK}=\delta_{kl} \qquad \bar{\chi}_{Kk}\bar{\chi}_{Lk}=\delta_{KL}.
\end{align}
It is straightforward to prove that
\begin{align}
    \chi_{kK}=\bar{\chi}_{Kk}.
\end{align}
Consequently, the righthand part of Eq.~\ref{eq:inverse} becomes
\begin{align}
    \chi_{kK}\chi_{kL}=\delta_{KL} .
\end{align}
Here and throughout, an index followed by a comma denotes a partial derivative, e.g.,
\begin{align}
    x_{k,K}=\frac{\partial x_k}{\partial X_K}  \qquad\mbox{and}\qquad
      X_{K,k}=\frac{\partial X_K}{\partial x_k}.
\end{align}

For fluid flow, deformation-rate tensors are used to characterize the
viscous resistance.  Deformation-rate tensors may be deduced by
calculating the material time derivative of the spatial deformation
tensors.  For the morphing continuum, two objective deformation-rate tensors
are
\begin{equation}
   a_{kl}=v^{\text{MCT}}_{l,k}+e_{lkm}\omega^{\text{MCT}}_m
    \qquad\mbox{and}\qquad   b_{kl}=\omega^{\text{MCT}}_{k,l},
\end{equation}
where $v^{\text{MCT}}_k$ is the velocity vector and
$\omega^{\text{MCT}}_k$ is the self-spinning gyration vector.  If a
fluid molecule is assumed to be a rigid body particle, it possesses two
types of motion, translational velocity ($v^{\text{MCT}}_k$), found
by solving the MCT linear momentum equation, and spinning gyration
($\omega^{\text{MCT}}_k$) found by solving the MCT angular momentum
equation.  In the classical Navier-Stokes equations, only the
translational velocity can be directly solved from the balance law of
linear momentum.  To investigate the effect of the rotational motion of
the molecule, one must use the velocity field and take the angular
velocity to be one-half of the vorticity i.e.,
$\frac{1}{2}e_{ijk}v^{\text{MCT}}_{j,i}$.  This approximation in the
Navier-Stokes equations limits not only predicting the flow physics
involving molecular spinning, but also fails to represent the
interaction between translation and spinning.

Thermodynamic balance laws for morphing continuum theory include
(1) conservation of mass; (2) conservation of linear momentum; (3)
conservation of angular momentum; (4) conservation of energy; and (5)
the Clausius-Duhem inequality.  All five can be expressed as follows:

{\textit {Balance Law of mass}}
\begin{align}
    \frac{\partial\rho}{\partial t}+\left(\rho v^{\text{MCT}}_{i,i}\right)=0
\label{eq:MCT+Mass}
\end{align}

{\textit {Balance Law of linear momentum}}
\begin{align}
    t_{lk,l}+\rho\left(f_k-\dot{v}^{\text{MCT}}_k\right)=0
\label{eq:MCT+LM}
\end{align}

{\textit {Balance Law of angular momentum}}
\begin{align}
    m_{lk,l}+e_{ijk}t_{ij}+\rho i_{km}\left(l_m-\dot{\omega}^{\text{MCT}}_m\right)=0
\label{eq:MCT+AM}
\end{align}

{\textit {Balance Law of energy}}
\begin{align}
    \rho\dot{e}-t_{kl}a_{kl}-m_{kl}b_{lk}+q_{k,k}=0
\label{eq:MCT+E}
\end{align}

{\textit {Clausius-Duhem inequality}}
\begin{align}
    \rho\left(\dot{\psi}+\eta\dot{\theta}\right)+t_{kl}a_{kl}+m_{kl}b_{lk}-\frac{q_k}{\theta}\theta_{,k}\geq0
\label{eq:CDI}
\end{align}
where $\rho$ is mass density, $i_{km}$ the microinertia for the shape of the
microstructure, $f_k$ the body force density, $l_m$ the body moment
density, $e$ the internal energy density, $\eta$ the entropy density,
$\psi=e-\theta\eta$ the Helmholtz free energy, $t_{lk}$ the Cauchy stress,
$m_{lk}$ the  moment of stress, and $q_k$ the heat flux.  It is worthwhile to
mention that $t_{lk}$, $m_{lk}$ and $q_k$ are the constitutive equations
for the morphing continuum theory and can be derived from the
Clausius-Duhem inequality (see Eq.~\ref{eq:CDI}) through the Coleman-Noll
procedure \cite{ChenLeeLiang2011, ColemanNoll1963}.

The concept of microinertia,
$i_{km}\equiv{\int\rho'\xi_k\xi_mdv'}/{\int\rho'dv'}\equiv\langle\xi_k\xi_m\rangle$,
is similar to the moment of inertia in rigid body rotation and measures the
resistance of the microstructure to changes to its rotation.  It can be
further expressed as
\begin{equation}
   j_{km}=i_{pp}\delta_{km}-i_{km} \qquad\mbox{where}\qquad
   j\equiv\frac{1}{3}j_{pp}.
\end{equation}

The volume $v'$ refers to the volume of the finite-size microstructure. 
If the finite-size particle is assumed to be a solid sphere with a radius
$d$ and a constant density $\rho$, the microinertia can be computed as
$j=\frac{2}{5}d^2$. This result shows the microinertia for a spherical
microstructure is the moment of inertia of a sphere divided by its mass.
The experimental data of Lagrangian velocities of a trace particle can
be used to determine the geometry of the microstructure
\cite{ChenLiangLee2011, MordantMetzMichelPinton2001}.

There are multiple different definitions for fluids, including (1)
fluids do not have a preferred shape \cite{Batchelor2000}, and (2), fluids
cannot withstand shearing forces, however small, without sustained
motion \cite{Panton2013}.  Nevertheless, all these definitions describe
the physics of fluid flow, and yet provide little help in mathematically
formulating a continuum theory for fluids.  In rational continuum
thermomechanics, Eringen \cite{Eringen1980} formally defined fluids by saying that
``a body is called fluid if every configuration of
the body leaving density unchanged can be taken as the reference
configuration."

This definition implies $x_{k,K}\rightarrow\delta_{kK}$ and
$\chi_{kK}\rightarrow\delta_{kK}$ where $\delta_{kK}$ is the shifter,
the directional cosine between the current configuration and reference
configuration.  Hence, the difference between the reference configuration and
the current configuration is just a coordinate transformation.  As a result,
the axiom of objectivity should always be considered and obeyed.
The axiom of objectivity, or frame-indifference, states that
the constitutive equations must be form-invariant with respect to rigid body motions of the spatial frame
of reference \cite{Eringen1971, Eringen1980}.

The state of fluids in morphing continuum theory is expressed by the
characterization of the response functions $\mathbf{Y}=\{\psi, \eta,
t_{kl}, m_{kl}, q_k\}$ as functions of a set of independent variables
$\mathbf{Z}=\{\rho^{-1}, \theta, \theta_{,k}, a_{kl}, b_{kl}\}$.  In
addition, following the axiom of equipresence, at the outset the
constitutive relations are written as
$\mathbf{Y}=\mathbf{Y}(\mathbf{Z})$.

The Clausius-Duhem inequality of Eq.~\ref{eq:CDI}, also known as the
thermodynamic second law, is a statement concerning the irreversibility
of natural processes, especially when energy dissipation is involved.
Feynman et.\ al.\ stated ``so we see that a substance must be limited in a
certain way; one cannot make up anything he wants; ...  This [entropy]
principle, this limitation, is the only rule that comes out of
thermodynamics."  \cite{FeymanLighttonSand1970}  After the Coleman-Noll
procedure, i.e., combining the inequality with the response function and
the independent variables, Eq.~\ref{eq:CDI} reduces to
\begin{align}
   t_{kl}^d a_{kl}+m_{kl} b_{lk}-\frac{q_k \theta_{,k}}{\theta}\geq0 .
\end{align}
There are three pairs of thermodynamic conjugates, ($t_{kl}^d$,
$a_{kl}$), ($m_{kl}$, $b_{lk}$), and
($\frac{q_k}{\theta}$,$\theta_{,k}$) that contribute to the
irreversibility of the material.  A set of the thermodynamic fluxes
$\mathbf{J}$ is defined as $\mathbf{J}=\{t_{kl}^d, m_{kl},
\frac{q_k}{\theta}\}$ and are functions of a set of the thermodynamic forces
($\mathbf{Z}^D$) and other independent variables ($\mathbf{Z}^R$),
$\mathbf{Z}=\{\mathbf{Z}^R; \mathbf{Z}^D\}=\{\rho^{-1}, \theta; a_{kl},
b_{lk}, \theta_{,k}\}$.  With these sets of thermodynamic fluxes and
thermodynamic forces, the Clausius-Duhem inequality can be rewritten as
\begin{align}
   \mathbf{J}\left(\mathbf{Z}^R; \mathbf{Z}^D\right)\cdot\mathbf{Z}^D\geq0 .
\end{align}

Onsager and Edelen proposed that the thermodynamic fluxes can be
obtained by the general dissipative function \cite{Chen2013, Edelen1972,
Onsager1931_1, Onsager1931_2}
\begin{align}
   \mathbf{J}=\frac{\partial
   \Psi\left(\mathbf{Z}^R,\mathbf{Z}^D\right)}{\partial\mathbf{Z}^D}+\mathbf{U} ,
\end{align}
where the vector-valued function $\mathbf{U}$ is the constitutive
residual with $\mathbf{Z}^D\cdot\mathbf{U}=0$.  This result indicates
that $\mathbf{U}$ does not contribute to the dissipative or entropy
production.  For simplicity, one can further set $\mathbf{U}=0$.

To determine thermodynamic fluxes for a fluid using the derivative of
$\Psi$ with respect to the thermodynamic forces $\mathbf{Z}^D$, $\Psi$ needs to be
invariant under superimposed rigid body motion, i.e., the dissipative
function $\Psi$ must satisfy the axiom of objectivity \cite{Chen2013}.
Hence, $\Psi$ is an isotropic function of scalar and can be expressed by
Wang's representation theorem \cite{Wang1969, Wang1970} as
\begin{equation}
   \Psi\{\mathbf{Z}^R, \mathbf{Z}^D\}=\Psi\{I_1, I_2, I_3, ..., I_n\}
   \qquad\mbox{and}\qquad
   \mathbf{J}=\frac{\partial\Psi}{\partial\mathbf{Z}^D}=
     \sum_{i=1}^{n}\frac{\partial\Psi}{\partial I_i}
     \frac{\partial I_i}{\partial \mathbf{Z}^D} .
\end{equation}
It should be noted here that $b_{lk}$ and $m_{kl}$ are pseudo-tensors
while the rest, including $\theta_{,k}$, $t_{kl}^d$, $q_k$ and
$a_{kl}$ are normal tensors \cite{ChenLeeLiang2011}.  Considering the
mixing of pseudo-tensors and normal tensors in $\mathbf{Z}^R$ and
$\mathbf{Z}^D$ for linear constitutive equations, the set of invariants
includes $I_1=a_{(ii)}$, $I_2=a_{(ij)}a_{(ji)}$, $I_3=b_{(ij)}b_{(ji)}$,
$I_4=\theta_{,k}\theta_{,k}$, $I_5=a_{[ij]}a_{[ji]}$, $I_6=b_{[ij]}b_{[ji]}$,
and $I_7=e_{ijk}b_{ij}\theta_k$ .
Here (...) refers to the symmetric part, [...] indicates the
anti-symmetric part and $e_{ijk}$ is the permutation symbol.  Hence, the
thermodynamic fluxes can be further derived as
\begin{align}
   &t_{kl}=t_{\left(kl\right)}+t_{\left[kl\right]}=\frac{\partial\Psi}{\partial
   a_{\left(kl\right)}}+\frac{\partial\Psi}{\partial
   a_{\left[kl\right]}}=\frac{\partial\Psi}{\partial
   I_1}\delta_{kl}+\frac{\partial\Psi}{\partial
   I_2}a_{\left(kl\right)}+\frac{\partial\Psi}{\partial I_5}a_{\left[kl\right]}\nonumber \\
   &m_{kl}=m_{\left(kl\right)}+m_{\left[kl\right]}=\frac{\partial\Psi}{\partial
   b_{\left(kl\right)}}+\frac{\partial\Psi}{\partial
   b_{\left[kl\right]}}=\frac{\partial\Psi}{\partial
   I_3}b_{\left(kl\right)}+\frac{\partial\Psi}{\partial
   I_6}b_{\left[kl\right]}+\frac{\partial\Psi}{\partial
   I_7}e_{klm}\theta_{,m}\nonumber \\
   &q_k=\frac{\partial\Psi}{\partial\theta_{,k}}=\frac{\partial\Psi}{\partial
   I_4}\theta_{,k}+\frac{\partial\Psi}{\partial I_7}e_{ijk}b_{ij}
\end{align}
These equations can also be put into matrix form as
\begin{align}
\begin{bmatrix}
   t_{(kl)} \\
   t_{[kl]} \\
   m_{(kl)} \\
   m_{[kl]} \\
   q_k
\end{bmatrix}
=
\begin{bmatrix}
   \lambda\delta_{kl}+2\mu+\kappa       &0      &0      &0      &0\\
   0    &\kappa &0      &0      &0\\
   0    &0      &\alpha\delta_{kl}+\frac{1}{2}\left(\beta+\gamma\right) &0      &0\\
   0    &0      &0      &\frac{1}{2}\left(\beta-\gamma\right)   &\frac{\alpha_Te_{klm}}{\theta}\\
   0    &0      &0      &\frac{\alpha_Te_{klm}}{\theta} &K
\end{bmatrix}
\begin{bmatrix}
   a_{\left(kl\right)}\\
   a_{\left[kl\right]}\\
   b_{\left(kl\right)}\\
   b_{\left[kl\right]}\\
  \theta_{,m}
\end{bmatrix} .
\label{eq:ThermoMatrix}
\end{align}
Notice the symmetry of this thermodynamic matrix.
Equations \ref{eq:ThermoMatrix} connect the thermodynamic fluxes and
the thermodynamic forces, and can be referred to as Onsager's reciprocal
relations derived in 1931 \cite{Onsager1931_1, Onsager1931_2} leading
to his Nobel Prize in Chemistry in 1968.  With further algebraic
manipulation, the linear constitutive equations for the morphing continuum
are
\begin{align}
   &t_{kl}=-p\delta_{kl}+\lambda v^{\text{MCT}}_{m,m}\delta_{kl}+\left(\mu+\kappa\right)
   \left(v^{\text{MCT}}_{l,k}+e_{lkm}\omega^{\text{MCT}}_m\right)+\mu\left(v^{\text{MCT}}_{k,l}+
   e_{klm}\omega^{\text{MCT}}_m\right)\nonumber .\\
   &m_{kl}=\frac{\alpha_T}{\theta}e_{klm}\theta_{,m}+\alpha\omega^{\text{MCT}}_{m,m}
   \delta_{kl}+\beta\omega^{\text{MCT}}_{k,l}+\gamma\omega^{\text{MCT}}_{l,k}\nonumber\\
   &q_k=\frac{\alpha_T}{\theta}e_{klm}\omega^{\text{MCT}}_{k,l}+K\theta_{,m}
\label{eq:MCT+Con}
\end{align}

\bibliography{mybibfile}

\begin{thebibliography}{10}
\expandafter\ifx\csname url\endcsname\relax
  \def\url#1{\texttt{#1}}\fi
\expandafter\ifx\csname urlprefix\endcsname\relax\def\urlprefix{URL }\fi
\expandafter\ifx\csname href\endcsname\relax
  \def\href#1#2{#2} \def\path#1{#1}\fi

\bibitem{Huang1963}
K.~Huang, Statistical Mechanics, John Wiley \& Sons, New York, NY, 1963.

\bibitem{Eringen1971}
A.~C. Eringen, Continuum Physics, Academic Press, New York, NY, 1971.

\bibitem{Eringen1980}
A.~C. Eringen, Mechanics of Continua, Robert E. Krieger, Huntington, NY, 1980.

\bibitem{Truesdell1966}
C.~Truesdell, Continuum Mechanics I: The Mechanical Functions of Elasticity and
  Fluid Dynamics, Science, New York, NY, 1966.

\bibitem{TruesdellRajagopal1999}
C.~Truesdell, K.~R. Rajapogal, An Introduction to the Mechanics of Fluids,
  Birkhauser, Boston, MA, 1999.

\bibitem{Chen2013}
J.~Chen, Extension of nonlinear onsager theory of irreversibility, Acta
  Mechanica 224 (2013) 3153--3158.

\bibitem{ChenLeeLiang2011}
J.~Chen, J.~D. Lee, C.~Liang, Constitutive equations of micropolar
  electromagnetic fluids, Journal of Non-Newtonian Fluid Mechanics 166 (2011)
  867--874.

\bibitem{CimmelliSellitto2009}
V.~A. Cimmelli, A.~Sellitto, V.~Trian, A generalized coleman{\textendash}noll
  procedure for the exploitation of the entropy principle, Proceedings of the
  Royal Society of London A: Mathematical, Physical and Engineering
  Sciences\href {http://dx.doi.org/10.1098/rspa.2009.0383}
  {\path{doi:10.1098/rspa.2009.0383}}.

\bibitem{Bowen1976}
R.~M. Bowen, Theory of mixture, in: A.~C. Eringen (Ed.), Continuum Physics,
  Academic Press, Waltham, 1976, pp. 11--32.

\bibitem{Eringen1997}
A.~C. Eringen, A mixture theory of electromagnetism and superconductivity,
  International Journal of Engineering Science 35 (1997) 1137--1157.

\bibitem{ColemanNoll1963}
B.~D. Coleman, W.~Noll, The thermodynamics of elastic materials with heat
  conduction and viscosity, Archive for Rational Mechanics and Analysis 13
  (1963) 167--178.

\bibitem{Truesdell1984}
C.~Truesdell, Rational Thermodynamics, Springer-Verlag, New York, NY, 1984.

\bibitem{Eringen1964_2}
A.~C. Eringen, Simple microfluids, International Journal of Engineering Science
  2 (1964) 205--217.

\bibitem{Eringen1964_1}
A.~C. Eringen, Theory of micropolar fluids, Journal of Applied Mathematics and
  Mechanics 16 (1964) 1--8.

\bibitem{Alizadeth2011}
M.~Alizadeth, G.~Silber, A.~G. Nejad, Continuum mechanical gradient theory with
  an application to fully developed turbulent flows, Journal of Dispersion
  Science and Technology 32 (2011) 185--192.

\bibitem{Ahmadi1975}
G.~Almadi, Turbulent shear flow of micropolar fluids, International Journal of
  Engineering Science 13 (1975) 959--964.

\bibitem{Peddieson1972}
J.~Peddieson, An application of the micropolar fluid model to the calculation
  of a turbulent shear flow, International Journal of Engineering Science 10
  (1972) 23--32.

\bibitem{Wonnell2017}
L.~Wonnell, J.~Chen, Morphing continuum theory: Incorporating the physics of
  microstructures to capture the transition to turbulence within a boundary
  layer, Journal of Fluids Engineering 139 (2017) 011205.

\bibitem{Chen2016}
J.~Chen, Advanced kinetic theory for polyatomic gases at equilibrium, AIAA
  paper.

\bibitem{Delhommelle2002}
J.~Delhommelle, D.~J. Evan, Poiseuille flow of a micropolar fluid, Molecular
  Physics 100 (2002) 2857--2865.

\bibitem{Papautsky1999}
I.~Papautsky, J.~Brazzle, T.~Ameel, A.~B. Frazier, Laminar fluid behavior in
  microchannel using micropolar fluid theory, Sensors and Actuators 73 (1999)
  101--108.

\bibitem{Boltzmann1872}
L.~Boltzmann, Weitere studien {\"u}ber das w{\"a}rmegleichgewicht unter
  gasmolek{\"u}len, Wiener Berichte 66 (1872) 275--370.

\bibitem{Curtiss1992}
C.~F. Curtiss, The classical boltzmann equation of a molecular gas, Journal of
  Chemical Physics 97 (1992) 1416.

\bibitem{Curtiss1981}
C.~F. Curtiss, The classical boltzmann equation of a gas of diatomic molecules,
  Journal of Chemical Physics 75 (1981) 376.

\bibitem{EuOhr2001}
B.~C. Eu, Y.~G. Ohr, Generalized hydrodynamics, bulk viscosity, and sound wave
  absorption and dispersion in dilute rigid molecular gases, Physics of Fluids
  13 (2001) 744.

\bibitem{GrmelaLafleur1998}
M.~Grmela, P.~G. Lafleur, Kinetic theory and hydrodynamics of rigid body
  fluids, Journal of Chemical Physics 109 (1998) 6956.

\bibitem{Myong2004}
R.~S. Myong, A generalized hydrodynamic computational model for rarefied and
  microscale diatomic gas flows, Journal of Computational Physics 195 (2004)
  655--676.

\bibitem{ArimanCakmak1967}
T.~Ariman, A.~S. Cakmak, Couple stresses in fluids, Physics of Fluids 10 (1967)
  24975.

\bibitem{Stokes1966}
V.~Stokes, Couple stresses in fluids, Physics of Fluids 9 (1966) 1709.

\bibitem{Klembt2015}
S.~Klembt, E.~Durupt, S.~Datta, T.~Klein, A.~Baas, Y.~Leger, C.~Kruse,
  Exciton-polariton gas as a nonequilibrium coolant, Physical Review Letters
  114 (2015) 186403.

\bibitem{Venugopal2015}
V.~Venugopal, S.~S. Girimaji, Unified gas kinetic scheme and direct simulation
  monte carlo computations of high-speed lid-driven microcavity flows,
  Communications in Computational Physics 17 (2015) 1127--1150.

\bibitem{Wonnell2016}
L.~Wonnell, J.~Chen, Morphing continuum theory: Incorporating the physics of
  microstructures to capture the transition to turbulence within a boundary
  layer, AIAA paper.

\bibitem{Norouzi2015}
A.~Norouzi, E.~J. A., Two relaxation time lattice boltzmann equation for high
  knudsen number flows using wall function approach, Microfluidics and
  Nanofluidics 18 (2015) 323--332.

\bibitem{ChenLiangLee2011}
J.~Chen, C.~Liang, J.~D. Lee, Theory and simulation of micropolar fluid
  dynamics, Journal of Nanoengineering and Nanosystems 224 (2011) 31--39.

\bibitem{MordantMetzMichelPinton2001}
N.~Mordant, P.~Metz, O.~Michel, J.-F. Pinton, Measurement of lagrangian
  velocity in fully developed turbulence, Physical Review Letters 87 (2001)
  214501.

\bibitem{Batchelor2000}
G.~K. Batchelor, An Introduction to Fluid Dynamics, Cambridge University Press,
  New York, NY, 2000.

\bibitem{Panton2013}
R.~L. Panton, Incompressible Flow, Wiley, Hoboken, NJ, 2013.

\bibitem{FeymanLighttonSand1970}
R.~Feynman, R.~Leighton, M.~Sand, The Feynman Lectures on phsyics, Addison
  Wesley Longman, New York, NY, 1970.

\bibitem{Edelen1972}
D.~Edelen, A nonlinear onsager theory of irreversibility, International Journal
  of Engineering Science 10 (1972) 481--490.

\bibitem{Onsager1931_1}
L.~Onsager, Reciprocal relations in irreversible processes i, Physical Review
  37 (1931) 405--426.

\bibitem{Onsager1931_2}
L.~Onsager, Reciprocal relations in irreversible processes ii, Physical Review
  38 (1931) 2265--2279.

\bibitem{Wang1969}
C.-C. Wang, A new representation for isotropic functions, Archive for Rational
  Mechanics and Analysis 36 (1969) 166--223.

\bibitem{Wang1970}
C.-C. Wang, Corrigendum to representation for isotropic functions, Archive for
  Rational Mechanics and Analysis 43 (1970) 392--395.

\end{thebibliography}
\bibliographystyle{elsarticle-num}
\end{document}